\documentclass[12pt]{iopart}

\usepackage{graphicx}
\usepackage{amssymb}
\usepackage{subcaption}
\usepackage[abbr]{harvard}

\begin{document}

\title[Breast tissues in $\rho$-Z$_\textrm{eff}$   basis]{Characterization of breast tissues in density and effective atomic number basis via spectral X-ray computed tomography}

\author[cor1]{Stevan Vrbaški}
\address{Department of Physics, University of Trieste, Via Valerio 2, 34127 Trieste, Italy and Elettra-Sincrotrone Trieste S.C.p.A, 34149 Basovizza Trieste, Italy and INFN Division of Trieste, Via Valerio 2, 34127 Trieste, Italy}
\author{Lucia Mariel Arana Peña}
\address{Department of Physics, University of Trieste, Via Valerio 2, 34127 Trieste, Italy and Elettra-Sincrotrone Trieste S.C.p.A, 34149 Basovizza Trieste, Italy and INFN Division of Trieste, Via Valerio 2, 34127 Trieste, Italy}
\author{Luca Brombal}
\address{Department of Physics, University of Trieste, Via Valerio 2, 34127 Trieste, Italy and INFN Division of Trieste, Via Valerio 2, 34127 Trieste, Italy}
\author{Sandro Donato}
\address{Department of Physics, University of Calabria, Via Pietro Bucci, 87036 Arcavacata, Italy and INFN Division of Frascati, Via Enrico Fermi 54, 00044 Frascati, Italy}
\author{Angelo Taibi}
\address{Department of Physics and Earth Sciences, University of Ferrara, Via Saragat 1, 44122 Ferrara, Italy and INFN Division of Ferrara, Via Saragat 1, 44122 Ferrara, Italy}
\author{Adriano Contillo}
\address{Elettra-Sincrotrone Trieste S.C.p.A, 34149 Basovizza Trieste, Italy}
\ead{adriano.contillo@elettra.eu}
\author{Renata Longo}
\address{Department of Physics, University of Trieste, Via Valerio 2, 34127 Trieste, Italy and INFN Division of Trieste, Via Valerio 2, 34127 Trieste, Italy}
\begin{abstract}
\emph{Objective.} Differentiation of breast tissues is challenging in X-ray imaging because tissues might share similar or even the same linear attenuation coefficients $\mu$. Spectral computed tomography (CT) allows for more quantitative characterization in terms of tissue density and effective atomic number by exploiting the energy dependence of $\mu$. \emph{Approach.} In this work, 5 mastectomy samples and a phantom with inserts mimicking breast soft tissues were evaluated in a retrospective study. The samples were imaged at three monochromatic energy levels in the range of 24 - 38 keV at 5 mGy per scan using a propagation-based phase-contrast setup at SYRMEP beamline at the Italian national synchrotron Elettra. \emph{Main results.} A custom-made algorithm incorporating CT reconstructions of an arbitrary number of spectral energy channels was developed to extract the density and effective atomic number of adipose, fibro-glandular, pure glandular, tumor, and skin from regions selected by a radiologist. \emph{Significance.} Preliminary results suggest that, via spectral CT, it is possible to enhance tissue differentiation. It was found that adipose, fibro-glandular and tumorous tissues have average effective atomic numbers (5.94 $\pm$ 0.09, 7.03 $\pm$ 0.012, and 7.40 $\pm$ 0.10) and densities (0.90 $\pm$ 0.02, 0.96 $\pm$ 0.02, and 1.07 $\pm$ 0.03 g/cm$^{3}$) and can be better distinguished if both quantitative values are observed together.
\end{abstract}

\vspace{2pc}
\noindent{\it Keywords}: spectral breast computed tomography, material density, effective atomic number, photon-counting detectors, synchrotron radiation X-ray imaging 

\submitto{\PMB}

\section{Introduction}
Tissue differentiation in mammography is a challenging task due to the overlap of tissues with similar linear attenuation coefficients $\mu$ in the diagnostic energy range. A comparison of the attenuation of glandular tissue and close-by malignant tumors found that the average attenuation values are almost identical even in the low-energy range \cite{chen_measurement_2010,fredenberg_measurement_2018}. Computed tomography (CT) solves tissue overlapping issues and provides additional diagnostic information based on tissue morphology in 3D volumes. On the other hand, breast-dedicated CT uses higher X-ray energies compared to mammography due to the imaging of uncompressed breasts, thus bringing a further reduction in attenuation contrast between similar tissues. Tissue separation can be significantly improved using spectral CT imaging \cite{alvarez_energy-selective_1976}. New generations of whole-body spectral CT systems obtain source-generated spectral information from two (or more) scans at different energy levels (dual-energy), or detector-generated spectral response with layered and photon-counting (PC) detectors. The first breast CT scanners with PC detectors are being introduced in clinics \cite{kalender_technical_2017,berger_dedicated_2019,zellweger_breast_2022,schmidt_clinical_2022} and synchrotron facilities \cite{longo_towards_2016,longo_advancements_2019}, where, in the latter case, tomography can be performed with low-energy (20-40 keV) monochromatic beams. Spectral CT imaging can improve tissue separation by acquiring data at multiple energy levels, allowing for spectral material decomposition analysis and the creation of e.g., virtual monochromatic images, iodine maps, and density-effective atomic number ($\rho$/Z$_{\textrm{\tiny eff}}$) maps. In synchrotron facilities, source-generated spectral separation can be obtained using multiple
X-ray beams of precisely defined energies. Alternatively, when dealing with polychromatic spectra such as in clinical scanners, detectors with multiple thresholds (up to 12 \cite{danielsson_photon-counting_2021}) could be used, but the benefits of such an approach in breast CT imaging have not been explored yet. 

Given two (or more) scans at different energies, their content can be described as a linear combination of the attenuation of two basis materials where the weights of the combination are related to some physical property (e.g., volume fractions) of the basis materials. Two-basis decomposition is the most common approach because two physical effects contribute to image formation in the diagnostic energy range: the photoelectric effect and Compton scattering. Photoelectric-Compton basis itself is very convenient because of its physical relationship to density and effective atomic number \cite{alvarez_energy-selective_1976}, but decomposing biological materials in a pair of physical materials, for example, poly(methyl methacrylate) (PMMA) and aluminum (Al), leads to significantly lower decomposition errors \cite{champley_method_2019}, thus PMMA-Al basis was used in this study. Estimated coefficients for the linear combination of the two basis materials and the known energy dependence of their attenuation coefficients are used to extrapolate virtual monochromatic images (VMIs) at an arbitrary energy value within the diagnostic energy range \cite{mccollough_dual-_2015}. Clinical case studies show that virtual monochromatic images are diagnostically valuable for certain tasks \cite{albrecht_review_2019}, and they are a standard output in the first clinical whole-body PC-CT (Naeotom Alpha, Siemens) \cite{rajendran_first_2021}. Basis material maps are also used for the physical description of scanned objects in terms of material density $\rho$ and effective atomic number $Z_{\textrm{\tiny eff}}$. Several works have been published exploring this approach \cite{torikoshi_electron_2003,heismann_density_2003,szczykutowicz_simple_2011,azevedo_system-independent_2016,champley_method_2019,busi_effective_2019,vrbaski_spectral_2022}.
In fact, while different materials can share the same $\mu$ value for a given energy range, the successful physical description offers its unique characterization. Recent work on a System Independent Rho Z (SIRZ) software showed a potential for $\rho$/$Z_{\textrm{\tiny eff}}$ decomposition with poly-chromatic beams in low-density plastic materials with an accuracy within 2.1 $\%$ \cite{champley_method_2019}. 

In the present paper, for the first time, a spectral study on breast CT images is developed to characterize tissues in terms of material density $\rho$ and effective atomic number $Z_{\textrm{\tiny eff}}$. The work investigates the feasibility and the potential diagnostic benefit of $\rho$/Z$_{\textrm{\tiny eff}}$ decomposition in a dedicated phantom and on 5 mastectomy samples in a retrospective study. We also show the convenience of such decomposition for computing virtual monochromatic $\mu$ values at desired energy, taking into account the known dependence of X-ray attenuation on density and atomic number. The overarching goal required the development of methodological procedures that have been organized into several parts. The paper contains a theoretical approach to material and $\rho$/Z$_{\textrm{\tiny eff}}$ decomposition in section \ref{sec:theor}, calibration and Z$_{\textrm{\tiny eff}}$ definition in section \ref{sec:sample}, a description of breast mastectomy samples in section \ref{sec:breast}, and proposed solution to the well-known problem of the decomposition noise~\cite{niu_tu-f-18a-02_2014,dong_combined_2014,mechlem_dictionary-based_2016} in section \ref{sec:algo} which also provides a detailed explanation of how all parts are practically joined together. CT scans were acquired using synchrotron X-ray beams at several energies. The data collection was performed using a high-resolution PC detector, as detailed in section \ref{sec:setup}.  

\section{Materials and Methods}\label{sec:mam}

\subsection{Theoretical model}\label{sec:theor}

The attenuation of X-rays is influenced by several physical effects, which are dependent on the energy and material properties of the attenuator. In the range of photon energies useful for medical CT imaging, the attenuation coefficient $\mu$ of given material of density $\rho$, the atomic number $Z$, and atomic mass $A$ are approximated as the sum of the two contributions
\begin{equation}\label{eq:mu}
 \mu(\rho,Z,A,E)=\frac{\rho\,Z}{A}\,\left[K_{1}\,Z^{n-1}\,f_P(E)+K_{2}\,f_C(E)\right]
\end{equation}
where the functions $f_P(E)$ and $f_C(E)$ encode the energy dependencies of the photoelectric and Compton effects, respectively, and K$_{1}$, K$_{2}$ and $n$ are constants. Using material decomposition, the attenuation coefficient $\mu$ of a given material is expressed as a linear combination of the (known) attenuation coefficients of a pair of basis materials, here labeled $\mu_1$ and $\mu_2$
\begin{equation}\label{lincomb}
 \mu=x_1\,\mu_1+x_2\,\mu_2
\end{equation}
where $x_1$ and $x_2$ are the coordinates of the material in the reference frame identified by the selected basis. In the first step of our method, we used equation (\ref{lincomb}) to determine the coefficients $x_1$ and $x_2$, which represent the (energy-independent) relative concentrations of each basis material. Because more than two spectral scans were available for each sample \cite{piai_quantitative_2019}, instead of using traditional matrix inversion \cite{zhang_image_2019}, $x_1$ and $x_2$ coefficients were calculated by using a least-square fit of the form: 
\begin{equation}
 \sum_i\,\left(\mu(E_i)-(x_1\,\mu_1(E_i)+x_2\,\mu_2(E_i))\right)^2
\end{equation}
The fit procedure consists of a voxel-by-voxel minimization of the sum over the energies $E_i$ of the squared residuals. From such a point of view, including more images of different energies adds further points to the plane, which is expected to increase the accuracy of the decomposition.

Assuming $A=2Z$, as it is almost true for any chemical element with $Z\lesssim20$ and making use of equation (\ref{eq:mu}), the linear combination coefficients of equation~(\ref{lincomb}) read
\numparts
\begin{eqnarray}
  x_1 &= \frac{\rho\,Z_1\left(Z^nZ_2-Z\,Z_2^n\right)}{\rho_1Z\left(Z_1^nZ_2-Z_1Z_2^n\right)} \label{x1coord}\\
  x_2 &= \frac{\rho\,Z_2\left(Z\,Z_1^n-Z^nZ_1\right)}{\rho_2Z\left(Z_1^nZ_2-Z_1Z_2^n\right)} \label{x2coord} 
\end{eqnarray}
\endnumparts

$\rho_j$ and $Z_j$ being the density and the atomic number of the $j$-th basis material. 
It is straightforward to notice that equations~(\ref{x1coord}) and (\ref{x2coord}) depend on both the density and the atomic number of the material. In order to decouple the two dependencies, it is necessary to rotate the reference frame (by an angle $\phi=\arctan\!\left(\frac{\rho_2}{\rho_1}\right)$) and then rescale the second coordinate dividing it by the first one. The resulting coordinates, which read
\numparts
\begin{eqnarray}
x_\rho &= \frac{\rho}{\sqrt{\rho_1^2+\rho_2^2}}\label{xrhocoord}\\
x_Z &= \frac{Z^\ell\left(\rho_1^2+\rho_2^2\right)-\left(Z_1^\ell\rho_1^2+Z_2^\ell\rho_2^2\right)}{(Z_2^\ell-Z_1^\ell)\rho_1\rho_2}\label{xZcoord}
\end{eqnarray}
\endnumparts
(with $\ell=n-1$) are labeled $x_\rho$ and $x_Z$ because of their exclusive dependencies on the variables mentioned in the subscripts. Equations~(\ref{xrhocoord}) and (\ref{xZcoord}) represent the expected relationships between $x_\rho$ and $\rho$, and between $x_Z$ and $Z$, for a given choice of basis materials. Such expressions are used in the calibration procedure described in section~\ref{sec:phantomresults}, where the (known) densities and atomic numbers of reference materials will be fitted against the corresponding (measured) values of $x_\rho$ and $x_Z$ using the functional forms
\numparts
\begin{eqnarray}
x_\rho(\rho) &= \kappa\,\rho\label{xrhofunc}\\
x_Z(Z) &= p\,Z^\lambda+q\label{xZfunc}
\end{eqnarray}
\endnumparts
stemming directly from equations~(\ref{xrhocoord}) and (\ref{xZcoord}) with the (physically motivated) coefficients replaced by the effective fit parameters $\kappa$,  $\lambda$, $p$, and $q$. The resulting calibrated relationships map the measured $x_\rho$ and $x_Z$ of any imaged material onto its actual values of density and effective atomic number.

\subsection{Scan setup}\label{sec:setup}

The experimental image acquisition was carried out at the SYRMEP beamline of Elettra, Italian synchrotron light source in Trieste, in the framework of SYRMA-3D (SYnchrotron Radiation for MAmmography) collaboration \cite{longo_advancements_2019}. The SYRMEP beamline utilizes a laminar beam having a cross-section at the detector equal to 148.5 mm (horizontal) $\times$ 3.25 mm (vertical), while the energy is selected through a Si double-crystal (1,1,1) monochromator with a resolution around 0.1\%. To perform tomographic acquisition the sample was positioned on a rotation stage spinning at a constant speed of 4.5 degrees/s while 1200 projections were acquired over 180 degrees. The imaging detector was a large area high-resolution CdTe photon-counting device (Pixirad8) featuring a honeycomb matrix of 4096 $\times$ 476 pixels with a 60 $\mu$m pitch \cite{bellazzini_chromatic_2013,delogu_characterization_2016}. It was positioned $\sim$1.6 m from the sample to employ the propagation-based phase-contrast imaging technique. Acquired projections were pre-processed with an ad-hoc procedure \cite{brombal_large-area_2018} and then phase-retrieved using an algorithm based on the homogeneous transport of intensity equation (TIE-Hom)~\cite{paganin_simultaneous_2002}. Finally, CT reconstructions were obtained via a GPU-based filtered-back-projection algorithm with Shepp-Logan filtering \cite{brun_syrmep_2017}. It is worth noting that, despite being a product of phase retrieval, reconstructions are maps of the attenuation coefficients $\mu$ at a given energy, as thoroughly explained in \cite{gureyev_unreasonable_2017,brombal_phase-contrast_2018,piai_quantitative_2019}.

\subsection{Calibration phantom and effective atomic number}\label{sec:sample}

A custom-made cylindrical phantom with a 10 cm diameter dedicated to calibration and quality control of the synchrotron breast CT system was used \cite{contillo_proposal_2018,donato2022optimization}. The phantom was filled with water and contained five inserts of polyethylene (PE), nylon (PA), PMMA, polyoxymethylene (POM), and polytetrafluoroethylene (PTFE) mimicking soft tissues of similar attenuation properties. The density of each material was known and the effective atomic number was computed from the material composition. The atomic number is a physical property of an element, but the same concept cannot be trivially defined in compounds. Across the literature \cite{champley_method_2019,azevedo_system-independent_2016,spiers_effective_1946,tsai_physics_1976,puumalainen_assessment_1977,un_direct-zeff_2014} several definitions have been proposed, suggesting that it depends not just on material properties, but also experimental conditions \cite{bonnin_concept_2014}. We included most of the published definitions in open-source GUI software "Z{\it comp}ARE" \cite{vrbaski_zcompare_2022}. Part of the software is built on top of the python library (xraylib 4.1.0 package, Python 3.10) of the xraylib database \cite{brunetti_library_2004,schoonjans_xraylib_2011} and can be used to compare approaches for computing atomic numbers of compounds from this database or list of user-defined materials can be provided. Comparison between the methods using this software is analyzed in Appendix B, showing that depending on the method chosen, $Z_{\textrm{\tiny eff}}$ number can take different values for the same compound. However, any choice of the method led to a unique material description in chosen basis. Quantity $x_{Z}$ derived in section \ref{sec:theor}, was simply calibrated to the desired definition of effective atomic number for a compound by putting $Z = Z_{\textrm{\tiny eff}}$ in equation (\ref{xZfunc}). In this paper, we used the approach by Champley {\it et al} \citeyear{champley_method_2019} which defines $Z_{\textrm{\tiny eff}}$ of a compound as a linear combination of two consecutive Z numbers such that the least square error between X-ray transmission of the compound and the transmission of a combination of the two elements is minimized. Compounds' brute formula, density, and $Z_{\textrm{\tiny eff}}$ are given in table \ref{Tab1}. The phantom served two purposes: \emph{i)} to obtain the $\rho$ and $Z_{\textrm{\tiny eff}}$ calibration curve from the decoupled set in equations (\ref{xrhocoord}) and (\ref{xZcoord}) and \emph{ii)} to validate the theoretical model against the ground truth.

\begin{table}[ht]
\centering
\caption{List of materials composing the phantom together with chemical formulas, effective atomic numbers, and material density values\protect\cite{schoonjans_xraylib_2011}.}\label{Tab1}
\footnotesize
\begin{tabular}{@{}lllllll}
\br
\textbf{Material} & \textbf{Water} & \textbf{PE} & \textbf{PA} & \textbf{PMMA} & \textbf{POM} & \textbf{PTFE}\\
\mr
Brute formula & $\textrm{H}_2\textrm{O}$ & $\textrm{C}_2\textrm{H}_4$ & $\textrm{C}_{12}\textrm{H}_{22}\textrm{N}_2\textrm{O}_2$ & $\textrm{C}_5\textrm{H}_8\textrm{O}_2$ & $\textrm{C}\textrm{H}_2\textrm{O}$ & $\textrm{C}_2\textrm{F}_4$\\
\mr
Effective $Z$ & 7.44 & 5.28 & 6.16 & 6.49 & 7.01 & 8.56\\
Density (g/cm$^3$) & 1.0 & 0.94 & 1.14 & 1.19 & 1.425 & 2.2\\
\br
\end{tabular}
\end{table}
\normalsize
\subsection{Breast mastectomy samples}\label{sec:breast}
\begin{figure}[ht!]
    \centering
    \includegraphics[width=0.8\textwidth]{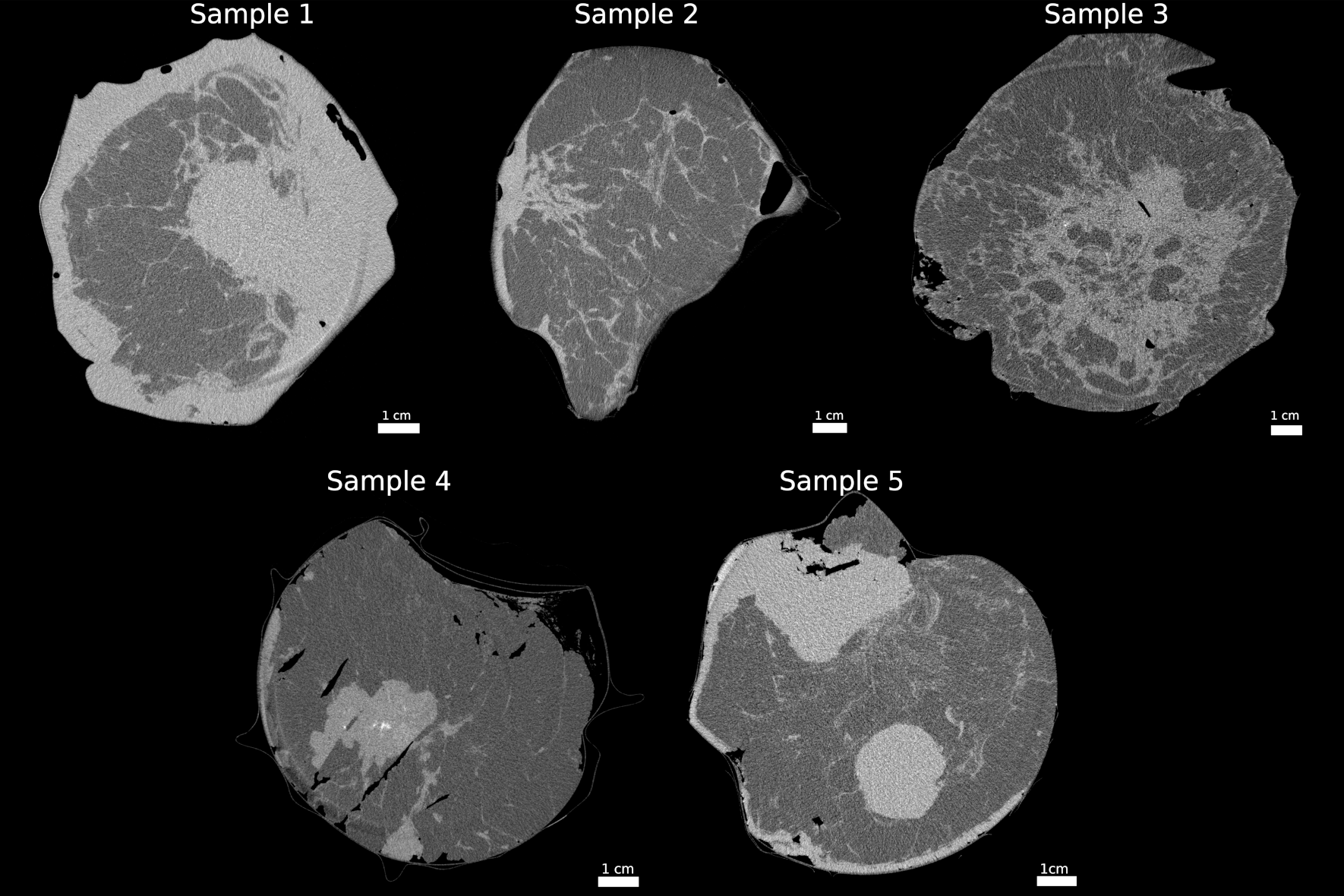}
    \caption{Mastectomy CT reconstructions acquired at SYRMEP beamline at Elettra Sincrotrone Trieste at the energy of 28 keV (Sample 1,2,3, and 5) and 26 keV (Sample 4).}
    \label{fig:breast}
\end{figure}

In addition to phantom scanning, post-mastectomy breast tissue images were analyzed in a retrospective study. The analyzed surgical samples were fixed in formalin, sealed in a vacuum bag, and conserved at 4 $^{\circ}$C. The preliminary analysis of the same data was published \cite{piai_quantitative_2019} as a feasibility study of the synchrotron breast CT approach. All the procedures adopted in this work followed Directive 2004/23/EC of the European Parliament and of the Council of 31 March 2004 on setting standards of quality and safety for the donation, procurement, testing, processing, preservation, storage, and distribution of human tissues. In the present work, we further processed the data to extract $\rho$ and $Z_{\textrm{\tiny eff}}$ of breast tissues. Tomographic reconstructions of selected samples are given in figure \ref{fig:breast}. They all contained adipose, fibro-glandular, and tumorous tissue, but only in sample 4 existed a region of glandular tissue clearly separated from the fibrous. It also contained calcification regions which were not evaluated in this study. All samples contained some type of malignant tissue: Samples 1, 2 and 3 contain infiltrating ductal carcinoma, Sample 4 contains infiltrating ductal carcinoma with a core of desmoplastic tissue, and Sample 5 contains vastly differentiated infiltrating ductal carcinoma. Samples 1,2,4, and 5 also contained portions of the skin.
The mean glandular dose of 5 mGy delivered per scan was computed according to a dedicated Geant4 Monte Carlo simulation \cite{fedon_geant4_2015,mettivier_glandular_2016}. The phantom scans were acquired at 25, 28, 32, and 35 keV, and mastectomy samples at several monochromatic beam energy levels in the range of 24 - 38 keV, from which three scans were selected for the spectral analysis.

\subsection{Practical implementation}\label{sec:algo}

\begin{figure}[ht!]
    \centering
    \includegraphics[scale=0.7]{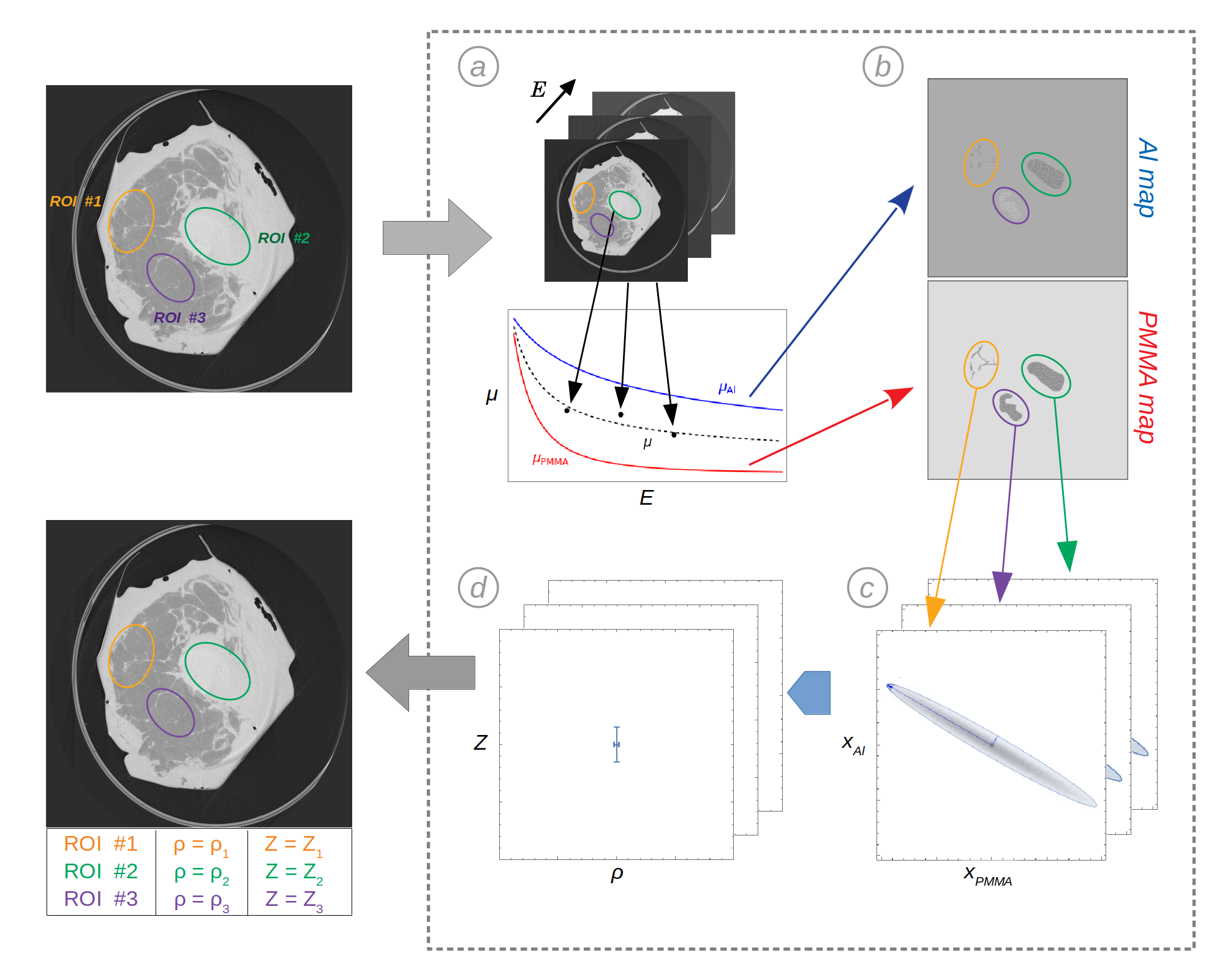}
    \caption{The step-by-step scheme of quantitative material evaluation followed in this work. Inputs to the algorithm (dashed rectangle) are delineated tissues. Section \emph{a} shows the material decomposition task which leads to material maps in section \emph{b}. These maps are represented in form of a 2D histogram and centers of clusters corresponding to each delineated material are extracted in section \emph{c}. These values are then transformed using equations~\ref{xrhocoord} and \ref{xZcoord} and an offline calibration using materials from table \ref{Tab1} is performed in section \emph{d} to obtain the $\rho$ and $Z_{\textrm{\tiny eff}}$ values as a final output.}
    \label{fig:scheme}
\end{figure}

The theoretical model described in section \ref{sec:theor} was implemented with Python (Python 3.10.0) and an interactive delineation tool was developed to select an arbitrary region of interest (ROI) within a reconstructed spectral data set (Matplotlib, Python 3.10.0) containing a single tissue type. In the first step of the process (framed with a dashed line in figure \ref{fig:scheme}), the voxel-to-voxel material decomposition method using the PMMA-Al basis is applied to the selected regions (figure \ref{fig:scheme}(a)), resulting in PMMA and Al maps for each ROI (figure \ref{fig:scheme}(b)). In the next step, the material maps are combined in 2D histograms, one for each tissue type (figure \ref{fig:scheme}(c)). Due to the presence of noise, the obtained histograms are blurred and elongated.
The correct decomposition coefficients $x_1$ and $x_2$ are considered to be the centers of obtained distributions. They are extracted using a 2D Gaussian fit method of the form:
\begin{equation}\label{gaussian}
G(x_1,x_2)= A\,e^{\left(-\frac{u^2}{2\,S_u^2}+\frac{v^2}{2\,S_v^2}\right)}
\end{equation}
$A$ is the intensity of the peak and $S_u$, $S_v$ the spreads (that is, the standard deviations of the associated distributions) along the major and minor axes, respectively. The peak coordinates $\bar{x}_1$ and $\bar{x}_2$ relative to $(x_1,x_2)$-plane of the histogram are contained in the quantities $u$ and $v$,
\numparts
\begin{eqnarray}
u = (x_1 - \bar{x}_1)\,\cos\theta + (x_2 - \bar{x}_2) \, \sin\theta\label{u}\\
v = (x_2 - \bar{x}_2) \,\cos\theta - (x_1 - \bar{x}_1) \, \sin\theta\label{v}
\end{eqnarray}
\endnumparts

$\theta$ is the tilt of the major axis of the spot with respect to the horizontal direction. After the fitting, the peak coordinates $\bar{x}_1$ and $\bar{x}_2$ found according to equations (\ref{u}) and (\ref{v}) are used in equations (\ref{x1coord}) and (\ref{x2coord}) to compute the final output of the procedure: the single values for densities and effective atomic numbers of selected tissues (figure \ref{fig:scheme}(d)). 
The measurement uncertainty, estimated as the standard error on the center of fitted 2D Gaussian distribution, is propagated through all mathematical transformation steps and measurement calibration. For the sake of conciseness, the complete analysis is given in the \ref{sec:noise}. 

\subsection{Data analysis}\label{sec:stats}

In addition to the experimental data obtained using the calibration phantom, the procedure described in the theoretical model (section \ref{sec:theor}) was also applied to published $\mu$ ("true") values \cite{schoonjans_xraylib_2011} of phantom material inserts at the energy levels used in the experiment. The true data points are used to evaluate the accuracy of the model by calculating the percentage error between the experimental and the ground truth data points, as given in equation (\ref{eq:accuracy}).

\begin{equation}\label{eq:accuracy}
\%\,error= \frac{|\textrm{ground\,truth - measured}|}{\textrm{ground\,truth}}\times \, 100
\end{equation}

The segmentation of the breast CT reconstructions was performed by an experienced radiologist, who selected ROIs containing adipose, fibro-glandular, glandular, skin, and tumorous tissue already knowing the mastectomy content from from specimen sampled for the histological examination. For each sample and tissue type, density and Z$_{\textrm{\tiny eff}}$ values and their uncertainties were estimated respectively as the mean and standard deviation evaluated over 10 consecutive CT slices. To estimate the discrimination power of $\rho$/Z$_{\textrm{\tiny eff}}$ maps, hence diagnostic potential, a mean Silhouette score using euclidean distance as a metric was computed on mean values of all tissue types collected from the mastectomy samples \cite{rousseeuw_silhouettes_1987}. Mean Silhouette score (MSS) is a tool to quantify how similar are samples within the same cluster and how separated they are from other clusters, defined as:
\begin{equation}\label{eq:mss}
\textrm{MSS} = \frac{1}{n} \sum_{k=0}^n\frac{\textrm{NC}_{k} - \textrm{IC}_{k}}{max(\textrm{NC}_{k},\,\textrm{IC}_{k})}
\end{equation}
where IC is the mean intra-cluster distance and NC is the mean nearest-cluster distance for each sample $k$, and $n$ equals number of selected tissue types $\times$ number of selected mastectomy samples. Negative values of MSS indicate cluster overlap, values around 0 signal that samples are on or close to the boundary between clusters, and positive values up to 1 indicate increased cluster separation. 

Finally, extracted density and effective atomic numbers from the calibration phantom and Sample 4 were fed to the mathematical relationship given in equation (\ref{eq:mu}) to extrapolate linear attenuation coefficients of tissues for arbitrary (virtual) monochromatic energy levels. The uncertainty was propagated from the standard deviation of obtained $\rho$ and Z$_{\textrm{\tiny eff}}$ values.

\section{Results}\label{sec:results}

\subsection{Calibration phantom results}\label{sec:phantomresults}

\begin{figure}[ht]
\centering
 \begin{subfigure}{0.47\textwidth}
  \centering
  \includegraphics[width=\linewidth]{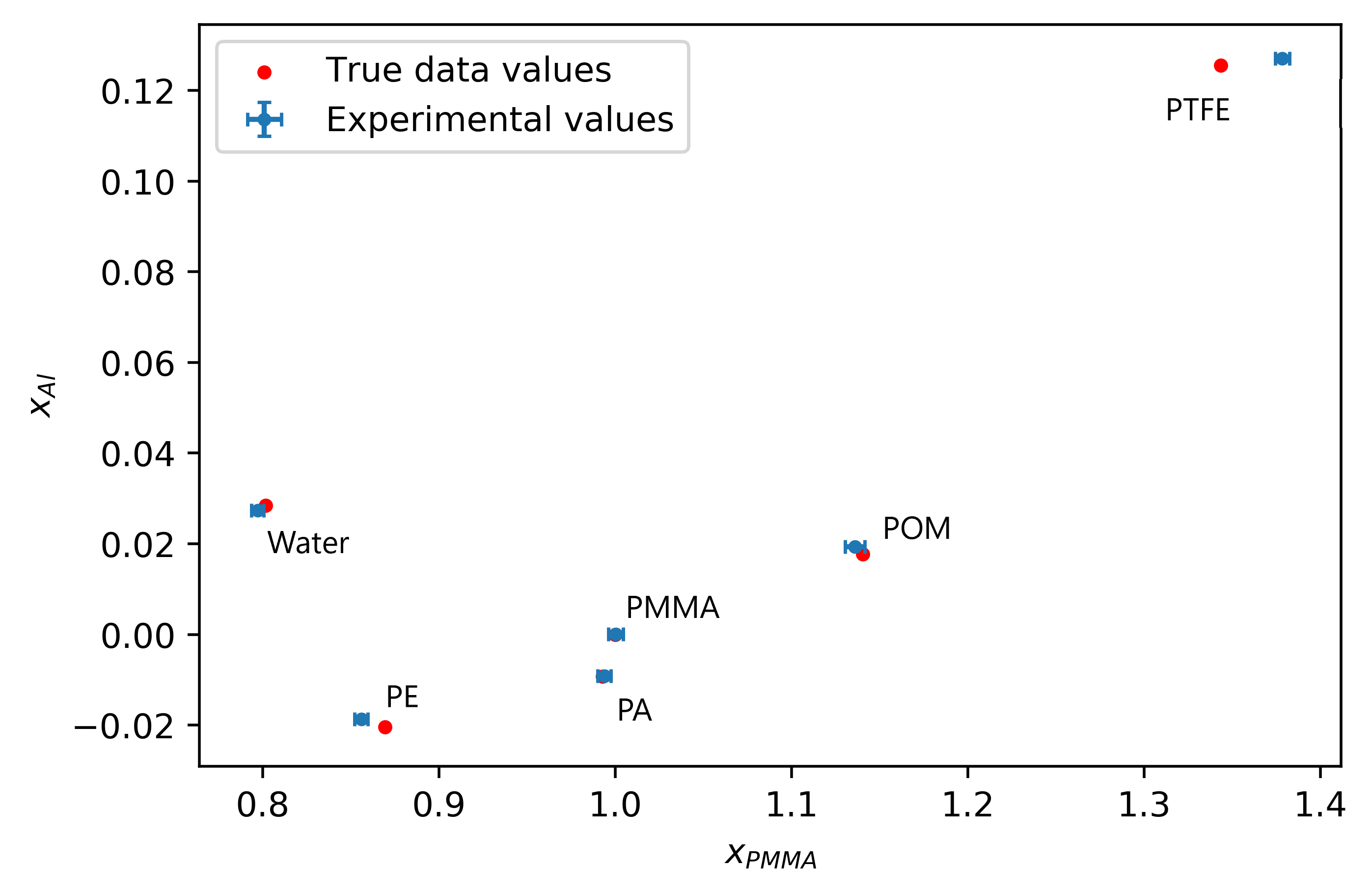}
  \caption{}\label{fig:positions}
 \end{subfigure}
 \quad
 \begin{subfigure}{0.47\textwidth}
  \centering
  \includegraphics[width=\linewidth]{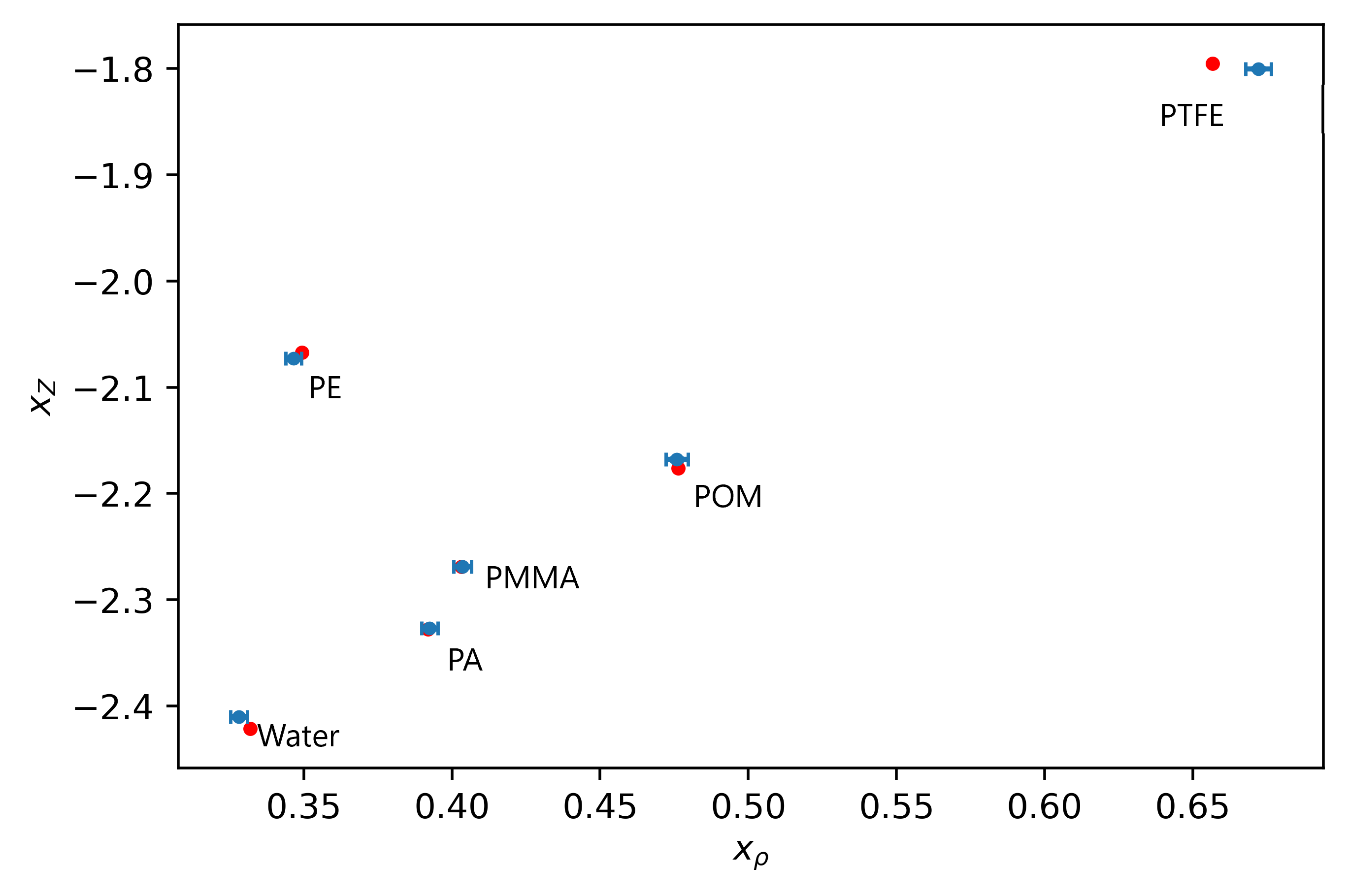}
  \caption{}\label{fig:xZxRho}
 \end{subfigure}
\caption{Experimentally obtained basis material concentrations (blue) against the true data values (red) in a) the PMMA-Al basis and b) decoupled reference frame. The error bars show standard error a) computed using equations (\ref{stderru}) and (\ref{stderrv}) and b) propagated using equations (\ref{sigmaxz}) and (\ref{sigmaxrho}) in \ref{sec:noise}.}
\label{fig:PAframe}
\end{figure}

The accuracy of material decomposition in phantom materials following the Gaussian fitting in the histogram space is shown in figure~\ref{fig:positions}. Quantities $x_1$ and $x_2$ in equations~(\ref{x1coord}) and (\ref{x2coord}) are replaced with $x_{\textrm{\tiny \textrm{PMMA}}}$ and $x_{\textrm{\tiny Al}}$ because of the particular basis choice. Figure~\ref{fig:xZxRho} corresponds to the decoupling of density and effective atomic number in equations~(\ref{xrhocoord}) and (\ref{xZcoord}). The fitting parameters $A$, $S_u$, $S_v$ defined in equation (\ref{gaussian}), and angle $\theta$ for phantom materials are given and further discussed in \ref{sec:gaussiannoise}.

\begin{figure}[ht]
\centering
 \begin{subfigure}{0.46\textwidth}
  \centering
  \includegraphics[width=\linewidth]{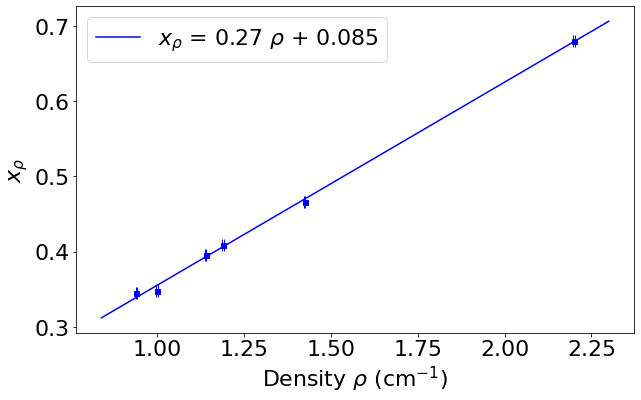}
  \caption{}\label{fig:cal_rho}
 \end{subfigure}
 \quad
 \begin{subfigure}{0.47\textwidth}
  \centering
  \includegraphics[width=\linewidth]{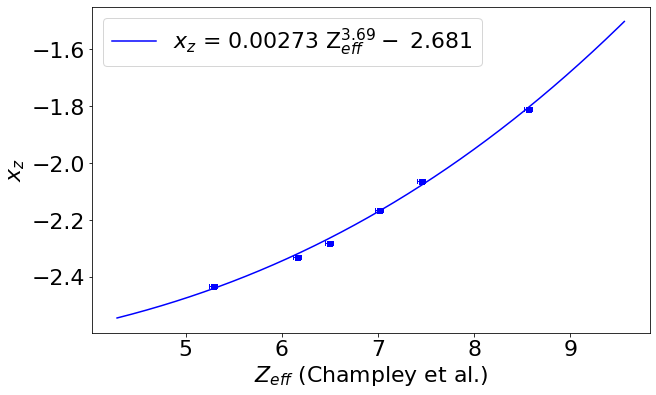}
  \caption{}\label{fig:cal_Z}
 \end{subfigure}
\caption{Theoretical data points fitted to (a) a linear calibration curve for density $\rho$, and (b) a power-law curve for $Z_{\textrm{\tiny eff}}$ of the same form suggested by the equations~(\ref{xrhofunc}) and (\ref{xZfunc}).}
\label{fig:cal_curves}
\end{figure}

The decoupled values from figure~\ref{fig:xZxRho} were calibrated to absolute density and effective atomic numbers (table \ref{Tab1}) applying the least-squares fitting method to the functional forms given in equations (\ref{xrhofunc}) and (\ref{xZfunc}). Both calibration curves were based on the phantom material inserts in the range of interest for soft tissue imaging and the mapping to correct density and effective atomic number was obtained at $R^2 =0.998$ and $R^2 =0.997$, respectively. Calibration curves are given in figure~\ref{fig:cal_curves} and corresponding density and $Z_{\textrm{\tiny eff}}$ values are given in table~\ref{Tab3}.
Using the equation \ref{eq:accuracy}, $\%$ errors in calibration materials were estimated to be below 3$\%$ and 1.5$\%$, respectively, after the density and effective atomic number calibration. 

\begin{table}[ht!] 
\centering
\caption{$Z_{\textrm{\tiny eff}}$ and $\rho$ with the corresponding standard errors $\sigma_{Z}$ and $\sigma_{\rho}$ obtained from the calibration phantom.\label{Tab3}}
\footnotesize
\begin{tabular}{@{}lllll}
\br
\textbf{Material} &  $\rho \pm \sigma_{\rho}\; \textrm{(g/cm}^3)$ & Z$_{\textrm{eff}}$ $\pm \sigma_{Z_{\textrm{\tiny eff}}}$ & $\% err\, \rho$ & $\% err\, \textrm{Z}_{\textrm{eff}}$\\
\mr
PE    & $0.963 \pm 0.004$ & $5.36 \pm 0.02$ & $2.4$ & $1.3$\\
Water & $0.971 \pm 0.001$ & $7.51 \pm 0.02$ & $2.9$ & $0.9$\\
PA    & $1.149 \pm 0.004$ & $6.10 \pm 0.01$ & $0.7$ & $1.0$\\
PMMA  & $1.198 \pm 0.004$ & $6.40 \pm 0.01$ & $0.7$ & $1.4$\\
POM   & $1.411 \pm 0.004$ & $7.04 \pm 0.01$ & $1.0$ & $0.4$\\
PTFE  & $2.204 \pm 0.003$ & $8.55 \pm 0.01$ & $0.2$ & $0.2$\\
\br
\end{tabular}
\end{table}
\normalsize

\subsection{Breast mastectomy results}\label{sec:breastresults}

\begin{figure}[ht!]
    \centering
    \includegraphics[scale=0.35]{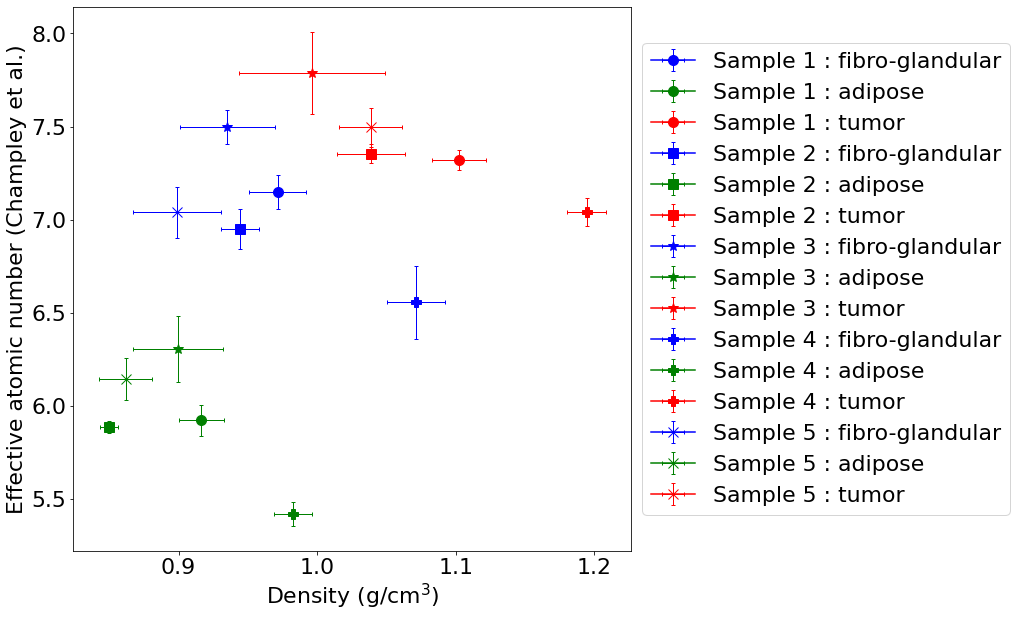}
    \caption{Quantitative description of tissues in terms of density (g/cm$^{3}$) and atomic number computed as in Champley {\it et al} \protect\citeyear{champley_method_2019}. Adipose, fibro-glandular and tumorous tissues are distinguished with green, blue, and red color, respectively, while marker shapes correspond to different mastectomy samples. Error bars represent the standard deviation of 10 ROIs.}
    \label{fig:sep}
\end{figure}

Tissues inside the mastectomies delineated by a radiologist were quantitatively analyzed in terms of their density and effective atomic number and presented in figure \ref{fig:sep}, with average values given in table \ref{Tab2a}. 
\begin{table}[ht!] 
\centering
\caption{Average $Z_{\textrm{\tiny eff}}$ and $\rho$ from 5 mastectomy samples.\label{Tab2a}}
\footnotesize
\begin{tabular}{@{}lll}
\br
\textbf{Tissue type} &  $\rho \pm SD_{\rho}\; \textrm{(g/cm}^3)$ & Z$_{\textrm{eff}}$ $\pm SD_{Z_{\textrm{\tiny eff}}}$ \\
\mr
Adipose         &  0.90 $\pm$ 0.02 & 5.94 $\pm$ 0.09\\
Fibro-glandular & 0.96 $\pm$ 0.02 & 7.03 $\pm$ 0.12\\
Tumor           & 1.07 $\pm$ 0.03 & 7.40 $\pm$ 0.10\\
\br
\end{tabular}
\end{table}
\normalsize

The level of separation between adipose, fibro-glandular and tumor tissue was found to be 0.31 (on a scale of -1 to 1) using MSS. Pairwise comparison of adipose and fibro-glandular (MSS = 0.59), adipose and tumor (MSS = 0.74), and fibro-glandular and tumor (MSS = 0.17) showed that adipose tissue can be well distinguished from the other two, while fibro-glandular and tumor are closer togerher in  $\rho$/$Z_{\textrm{\tiny eff}}$ space. From figure \ref{fig:sep} it can be seen that, adipose tissue can be distinguished from fibro-glandular and tumorous purely based on effective atomic number values. On the other hand, fibro-glandular and tumorous tissues are overlapping both in their effective atomic numbers and densities and can be well distinguished only if both quantitative values are observed together. 

In addition to adipose, fibro-glandular, and malignant tissue, the mastectomy labeled as Sample 4 contains well-separated pure glandular tissue and skin. Thus, an extensive investigation was conducted on this tissue. Extracted density and effective atomic number were  0.98 $\pm$ 0.01 and 5.42 $\pm$ 0.06 for adipose, 1.07 $\pm$ 0.02 and 6.56 $\pm$ 0.20 for fibro-glandular, 1.18 $\pm$ 0.02 and 6.88 $\pm$ 0.09 for glandular, 1.19 $\pm$ 0.01 and 7.04 $\pm$ 0.07 for tumorous, and 1.19 $\pm$ 0.02 and 6.87 $\pm$ 0.04 for skin tissue, respectively. In this particular mastectomy sample, mean density and $Z_{\textrm{\tiny eff}}$ are not distinguishable between the skin and glandular tissue. Glandular tissue has higher $Z_{\textrm{\tiny eff}}$ and density than fibro-glandular tissue, but almost the same density and slightly lower $Z_{\textrm{\tiny eff}}$ than tumorous tissue.

\begin{figure}[ht]
\captionsetup[subfigure]{labelformat=empty}
\centering
 \begin{subfigure}{0.48\textwidth}
  \centering
  \includegraphics[width=\linewidth]{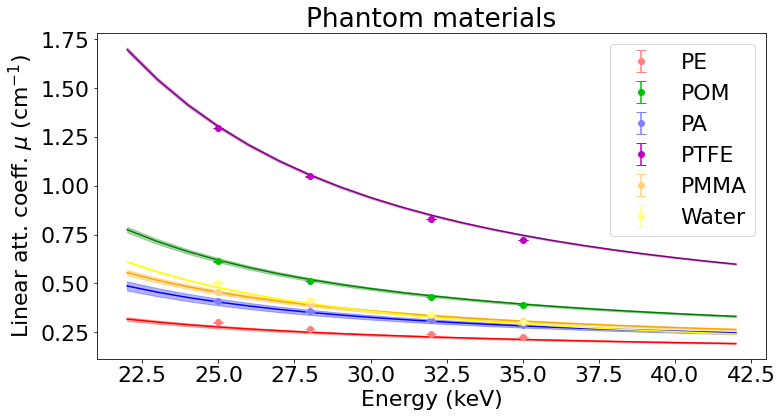}
  \caption{}
 \end{subfigure}
 \quad
 \begin{subfigure}{0.48\textwidth}
  \centering
  \includegraphics[width=\linewidth]{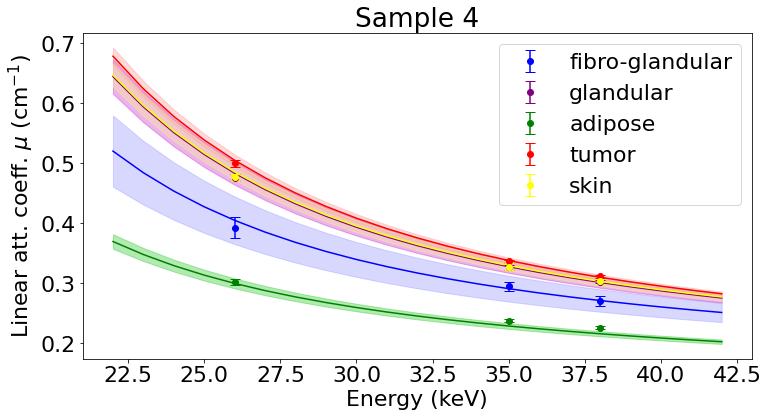}
  \caption{}
 \end{subfigure}
\caption{Virtual linear attenuation coefficients computed from extracted density and effective atomic numbers. Full lines represent virtual $\mu$ values extracted from experimental data at an arbitrary energy level, while uncertainty was propagated from the standard deviation of $\rho$ and $Z_{\textrm{\tiny eff}}$.}\label{fig:monos}
\end{figure}

Virtual $\mu$ values of phantom materials and tissues in sample 4 are given in figure \ref{fig:monos}. The obtained $\rho$ and $Z_{\textrm{\tiny eff}}$ values correctly map to experimentally measured $\mu$ values and provide information about tissue separation at lower energies with respect to those used for tissue scanning.

\section{Discussion}

Conventional CT scans offer a relatively low specificity when distinguishing between soft tissues of slightly different compositions. On the other hand, spectral imaging advances the ability to distinguish such tissues by probing their attenuation properties at several energy levels. Spectral information is used for the decomposition of data in basis materials, such as PMMA-Al. The decomposed material maps can be used as an intermediate step to estimating the uniquely defined physical quantities of imaged tissues. Utilizing these quantities, virtual monochromatic $\mu$ values at arbitrary energy levels can be further extrapolated. In this work, a new approach to material decomposition and $\rho$/$Z_{\textrm{\tiny eff}}$ estimation was presented to characterize breast tissues.

A least-square fitting approach to two-basis material decomposition was adopted for an over-determined system when $\mu$ values can be measured at several energy levels. This was the case in our retrospective study with monochromatic beams, but the same approach can be applied to multi-threshold photon counting detectors. The procedure for decoupling density from $Z_{\textrm{\tiny eff}}$ depends on the quality of performed material decomposition, but is at the same time independent of the method itself, and can be applied to other methods published in the literature. Experimentally computed concentrations $x_{\textrm{\tiny PMMA}}$ and $x_{\textrm{\tiny Al}}$ of basis materials and related basis $x_{\tiny \rho}$ and $x_{\textrm{\tiny Z}}$ in the phantom were found to be in good agreement with the ground truth values. Moreover, small associated standard errors in figure \ref{fig:PAframe} show that materials can be well separated. The theoretically derived functions in equations (\ref{xrhofunc}) and (\ref{xZfunc}) were fitted to 6 data points corresponding to tissue-equivalent plastics in the phantom. The high accuracy (R$^2 >$ 0.99) of the fitting procedure justifies the assumptions made in theoretical derivation. Density and $Z_{\textrm{\tiny eff}}$ values for phantom inserts, given in table \ref{Tab3}, showed excellent agreement with the ground truth data at an average $\%$ error of 1.34 $\%$ for $\rho$ and 0.89 $\%$ for $Z_{\textrm{\tiny eff}}$. These results demonstrate that soft-tissue-equivalent plastic materials of similar composition can be well distinguished from spectral CT data, based on their estimated effective atomic number and density value. 

The preliminary study on the breast cancer mastectomy samples was an attempt to demonstrate the feasibility of our method to distinguish between fibro-glandular and tumorous tissues inside the breast. Based on the available samples, it was shown that starting from spectral data it is possible to separate adipose, fibro-glandular and tumorous tissues based on their physical characteristics. This might be useful in risk assessment, cancer diagnosis and the assessment of the status of the disease. Although pure glandular and tumorous tissue in Sample 4 had almost the same density and slightly different effective atomic numbers, this could be due to the desmoplastic core present inside the tumor and no conclusions could be made based on single piece of evidence. The importance of $Z_{\textrm{\tiny eff}}$ to X-ray attenuation can be observed when comparing clusters related to adipose and fibro-glandular tissue. It can be seen that lower $\mu$ values for adipose tissue are driven by lower $Z_{\textrm{\tiny eff}}$, rather than significantly lower density. Distinguishing tumorous and pure glandular tissue is challenging because only slight differences exist in both density and $Z_{\textrm{\tiny eff}}$. In our study, we observed that it is not possible to distinguish between different tissues solely on density or effective atomic number or $\mu$ value alone while reasonable discrimination (MSS = 0.31) can be obtained considering 2D clusters in $\rho$/$Z_{\textrm{\tiny eff}}$ space. Therefore, using both density and effective atomic number maps in the diagnostic workflow could be beneficial, potentially allowing the identification of the tissue type based on quantitative measurements.

 Virtual monochromatic images are usually computed directly from decomposed basis (\emph{e.g.,} $\mu_1$ and $\mu_2$) using $x_1$ and $x_2$ in equation (\ref{lincomb}). Equivalently, physically relevant $\rho$/$Z_{\textrm{\tiny eff}}$ space can be used to compute other quantitative maps established in clinical practice. Linear attenuation values in figure \ref{fig:monos} were calculated using equation (\ref{eq:mu}) at energies within and outside the energy range used in the experiment. Virtual $\mu$ values calculated using the $Z_{\textrm{\tiny eff}}$ defined by Champley {\it et al} \citeyear{champley_method_2019} were in agreement (within the measurement error) with experimentally measured $\mu$ coefficients in both the phantom and Sample 4 and they enabled comparison with experimentally obtained $\mu$ values in other studies. For Sample 4, which contained tissues examined by Fredenberg {\it et al} \citeyear{fredenberg_measurement_2018}, $\mu$ values were compared at energies of 20, 30, and 40 keV with average \% error of 3.3, 7.3, and 3.4 \%, for glandular, adipose and tumor tissue, respectively.

Our approach to tissue analysis could be directly applied to state-of-the-art synchrotron radiation breast CT setups currently developed in Trieste at Elettra Sincrotrone SYRMEP beamline \cite{longo_advancements_2019}, and at ANSTO Imaging and Medical beamline in Australia \cite{tavakoli_taba_propagation-based_2021}. With these experimental settings, it should be mentioned that multiple energy acquisitions would required, thus resulting in an increased dose to the breast. On the other hand, due to the high contrast-to-noise ratio of phase-contrast images, the dose per scan could be reduced thus bringing overall acceptable radiation exposures. Considering clinical systems in hospitals, the advent of spectral CT paves the way to material decomposition following a single shot acquisition without a significant increment of the dose. Photon counting breast CT systems are in clinical practice \cite{berger_dedicated_2019} and an extension to spectral applications is expected.

Studies estimating both $\rho$ and $Z_{\textrm{\tiny eff}}$ of human tissues using synchrotron CT systems are almost nonexistent. The study by Torikoshi {\it et al} \citeyear{torikoshi_electron_2003} introduced a method to compute these quantities avoiding the material decomposition task. An average accuracy of 0.9 $\%$ for $\rho_{e}$ and 1 $\%$ for $Z_{\textrm{\tiny eff}}$ is comparable to our method, but the dose level used was not reported. The analysis was performed on low-$Z_{\textrm{\tiny eff}}$ plastic materials using a pair of monochromatic acquisitions. Considerably more papers using conventional systems have been published, but not focusing on breast tissues or breast dedicated scanners. Szczykutowicz {\it et al}~\citeyear{szczykutowicz_simple_2011} performed a methodologically similar approach to the one presented in this paper using a clinical scanner and test object without any noise remedying approach. They successfully decompose electron density and $Z_{\textrm{\tiny eff}}$ but at the cost of a significant reduction in signal-to-noise ratio. Lalonde {\it et al}~\citeyear{lalonde_general_2016} developed a model in which materials are decomposed in a compressed basis with principal component analysis, using the fact that human tissues are composed of a very limited number of elements. Then, the first principal components are virtual materials containing a certain fraction of those elements. From there $Z_{\textrm{\tiny eff}}$ were computed. Azevedo {\it et al}~\citeyear{azevedo_system-independent_2016} implemented their System-Independent-Rho-Z (SIRZ) method to obtain physical quantities of phantom materials independent of the shape of the X-ray spectrum. The photoelectric-Compton decomposition is performed in sinogram space and absolute $\rho_{e}$ and Z values are obtained after the calibration procedure. While physical characterization was successfully described for materials of fairly distinguishable compositions, the noise behavior was also not described in this work. Champley {\it et al} \citeyear{champley_method_2019} released a follow-up paper focusing on the optimization and simplification of spectral modeling as a new method called SIRZ-2. Most recently, Busi {\it et al}~\citeyear{busi_effective_2019} developed a physical characterization method using spectral detectors. They claim higher robustness and increased estimation accuracy ($~25\%$) compared to SIRZ methods. Extended work from the same group was published in~\cite{jumanazarov_significance_2021} to optimize the computation speed. Machine learning solutions to $\rho$/$Z_{\textrm{\tiny eff}}$ extraction were also tested by Su {\it et al}~\citeyear{su_machine_2018} using dedicated phantoms with several tissue-equivalent inserts. Good results in computing $Z_{\textrm{\tiny eff}}$ were obtained using artificial neural networks and the random forest method with the relative error between 1 and 2 $\%$ at clinically relevant doses. However, a low-dose scanning and evaluation of the model on materials that were removed from the training set led to errors up to 6 $\%$. Nonetheless, the authors showed that the machine learning approach is robust and computationally efficient. 

Considering specifically the estimation of $\rho$ and $Z_{\textrm{\tiny eff}}$ for breast tissues, only a few studies exist and have been made with experimental setups not used in diagnostic radiology. For instance, Gobo {\it et al} \citeyear{gobo_effective_2020} used a combination of transmission and scattering measurements with $^{241}$Am source and an X-ray tube, while Antoniassi {\it et al} \citeyear{antoniassi_study_2011} performed scattering measurements at 90 degrees by using low energy beams. Given diverse formulations of $Z_{\textrm{\tiny eff}}$ across the literature, we made a comparison in relative change using $Z_{\textrm{\tiny eff}}$ of nylon as a reference value in table \ref{Tab4}.
\begin{table} 
\centering
\caption{Literature review of experimentally obtained density and $Z_{\textrm{\tiny eff}}$ numbers for fibro-glandular, adipose and tumorous tissues.\label{Tab4}}
\footnotesize
\begin{tabular}{@{}cccccc}
\br
\textbf{Tissue type} & \multicolumn{2}{c}{\textbf{Density $\rho$ }} & \multicolumn{3}{c}{\textbf{$Z_{\textrm{\tiny eff}}$ to nylon \% diff}}\\
\mr
 & This work & Gobo et. al & This work & Gobo {\it et al} & Antoniassi {\it et al} \\
\mr
Fibro-glandular & 0.96 & 1.04 & 15.38 & 16.73 & 14.07\\
Adipose         & 0.90 & 0.95 & 2.68 & 4.40  & 5.60 \\
Tumorous        & 1.07 & 1.05 & 21.31 & 18.87 & 14.79\\
\br
\end{tabular}
\end{table}
\normalsize

Our study shows potential for quantitative breast imaging by translating spectral information into the computation of physically relevant quantities, but it also has some limitations. The accuracy of the presented method is mainly governed by the quality of material decomposition, which is highly dependent on the denoising of decomposed data. We gave up the spatial information inside the ROI to obtain quantitatively correct material decomposition during the 2D Gaussian denoising approach. Inaccurate tissue segmentation would result in an erroneous Gaussian fitting procedure as the number of peaks in 2D histogram space corresponds to the number of tissue types being evaluated at once. The appearance of a histogram containing several plastic materials was published in our previous study~\cite{vrbaski_quantitative_2021}. Thus, correct tissue segmentation is a critical aspect of the present model. For the proof of concept, we relied on high-quality synchrotron beam radiation, but to make this approach broadly used, we plan to apply the method to more accessible polychromatic sources. Finally, the conclusions drawn in this study were based on the analysis of 5 mastectomy samples. More samples will be evaluated to further confirm these findings. Despite the mentioned limitations, given the general validity of the proposed decomposition model and the foreseeable use of spectral detectors in breast CT scanners, the present feasibility study paves the way for its application to clinical spectral breast CT data.

\section{Conclusions}
A model incorporating CT reconstructions of an arbitrary number of spectral energy channels was developed to compute material density and effective atomic number. Density and effective atomic number of soft-tissue-equivalent plastic materials were computed with high accuracy, and the same approach was applied to the set of 5 mastectomy samples. Quantitative analysis suggests that adipose, fibro-glandular, and tumorous tissues could be well distinguished based on their effective atomic number and density. Density and effective atomic number can also be used for physics-based extrapolation of virtual monochromatic linear attenuation coefficients outside of the experimental energy range. 

\section{Acknowledgment}
The authors acknowledge the SYRMA-3D collaboration at Elettra for the support. We also would like to thank Dr. Cristina Marrocchio, Radiologist and Research fellow at the University of Parma, Italy supporting us in tissue identification from CT and histology data sets.

\appendix

\section{Noise analysis and uncertainties}\label{sec:noise}

As it was shown in section~\ref{sec:algo}, regions of uniform composition result in scattered values of elongated shape (\emph{i.e.} clusters) in the $(x_1,x_2)$-plane histogram. Such behavior was ascribed to the unavoidable amount of noise carried by the tomographic reconstructions on which the whole decomposition method is based. 
Detailed analysis of the propagation of measurement uncertainties, related to the size of the cluster, was carried out. The Gaussian fitting procedure introduced in section~\ref{sec:algo} individuates a preferred direction at angle $\theta$ aligned to the major axis $u$ of elliptical Gaussian. The length in this direction and the orthogonal one (minor axis) are identified as the Gaussian spreads in a reference frame $(x_u, x_v)$, which is parallel to the major and minor axes of the spot. The standard errors of the centroid coordinates can be approximated by the ratios between the Gaussian spreads $S_u$ and $S_v$ and the square root of the volume under the Gaussian surface
\numparts
\begin{eqnarray}
\sigma_{x_u}=\frac{S_u}{\sqrt{V}}\label{stderru}\\
\sigma_{x_v}=\frac{S_v}{\sqrt{V}}\label{stderrv}
\end{eqnarray}
\endnumparts
where the volume is given as
\begin{equation}
V = 2 \pi\,A\,\frac{S_u\,S_v}{b_1\,b_2}
\end{equation}
that it is equal to the total number of voxels contributing to the corresponding two-dimensional Gaussian function. Quantities $b_1$ and $b_2$ are bin sizes in both directions and $A$ is the peak value. Calculated errors are then translated into uncertainties of coordinates $x_\xi$ and $x_\zeta$ in the reference frame rotated by the angle $\phi$ defined in section~(\ref{sec:theor}). Rotation from the frame identified by angle $\theta$ to the frame identified by angle $\phi$ is defined as
\begin{equation*}
\left[\matrix{
x_\xi \cr
x_\zeta
}\right]
=
\left[\matrix{
\cos\alpha & \sin\alpha \cr
-\sin\alpha & \cos\alpha
}\right]\,
\left[\matrix{
x_u \cr
x_v
}\right]
\end{equation*}
where $\alpha = \phi - \theta$. Propagating uncertainty through the rotation of the reference frame leads to 
\numparts
\begin{eqnarray}
\sigma^2_{x_\xi} = \cos^2\alpha\, \sigma^2_{x_u} + \sin^2\alpha\, \sigma^2_{x_v}\\
\sigma^2_{x_\zeta} = \sin^2\alpha\, \sigma^2_{x_u} + \cos^2\alpha\, \sigma^2_{x_v}
\end{eqnarray}
\endnumparts
From the rotated frame, to the rescaled $x_{\rho},x_Z$ frame, the division of the second coordinate by the first one implicates

\begin{equation}\label{sigmaxz}
\sigma^2_{x_Z}= \frac{x^2_\zeta}{x^4_\xi}\sigma^2_{x_\xi}+\frac{1}{x^2_\xi}\sigma^2_{x_\zeta}
\end{equation}
while propagating $x_\rho$ remains trivial
\begin{equation}\label{sigmaxrho}
\sigma^2_{x_\rho}=\sigma^2_{x_1}.  
\end{equation}

Finally, shifting to the pair $\rho,Z$ requires equations~(\ref{xrhoZfuncs}), leading to the uncertainties
\numparts
\begin{eqnarray}
\sigma_\rho &= \kappa^{-1} \, \sigma_{x_\rho}\\
\sigma_Z &= \frac{(x_Z - q)^\frac{1-\lambda}{\lambda}}{\lambda\,p^{1/\lambda}}\,\sigma_{x_Z}\,.
\end{eqnarray}
\endnumparts
\section{Noise behaviour in 2D histogram space}\label{sec:gaussiannoise}

In addition to the estimated centers of the distributions in figure~\ref{fig:positions}, the 2D Gaussian fit method outputs several other parameters relevant for the accurate estimation of basis material concentrations summarized in table~\ref{Tab2}. The angle $\theta$ at which the cluster is extended remains constant ($\simeq-0.12$) for all materials indicating that blurring is not dependent on the material type. It is rather a result of combined contributions of image acquisition, reconstruction, and material decomposition noise. The amount of blurring in the major direction $\textrm{S}_{u}$ and the direction orthogonal to it $\textrm{S}_{v}$ is of the same order of magnitude for all materials. The $\textrm{S}_{u}$ values are around 2 orders of magnitude larger than $\textrm{S}_{v}$ values. The amplitude A of the Gaussian function remains constant for the same size of the ROI for all plastic inserts. 
\begin{table}[ht]
\caption{List of output parameters in 2D Gaussian fitting method for the phantom materials.\label{Tab2}}
\footnotesize
\begin{tabular}{@{}lllll}
\br
\textbf{Material} & \textbf{Amplitude (A)} &  \textbf{Major axis spread $\textrm{S}_{u}$} & \textbf{Minor axis spread $\textrm{S}_{v}$} & \textbf{Angle $\theta$}\\
\mr
Water & 38 & 0.40 & 0.0074 & -0.121\\
PE & 40 & 0.37 & 0.0065 & -0.121\\
PA & 38 & 0.32 & 0.0062 & -0.120\\
PMMA & 38 & 0.41 & 0.0073 & -0.121\\
POM & 38 & 0.32 & 0.0062 & -0.120\\
PTFE & 41 & 0.46 & 0.0078 & -0.121\\
\br
\end{tabular}
\end{table}
\normalsize
These properties are important for the Gaussian fitting procedure because an initial guess for fitting parameters can be given, improving the robustness of the method and increasing the computational speed. A more rigorous statistical description of these features will be the subject of a forthcoming standalone communication.

\section{Comparison of methods used to compute the atomic number of a compound - Z{\it comp}ARE}\label{sec:zcompare}

Effective atomic number as a property of a material depends on its chemical composition, but it is not uniquely defined in the literature. Very often different definitions are adopted by researchers depending on the experimental setup, energy range, and type of compounds. Several definitions that have been proposed to compute this quantity can be divided into: {\it i)} methods that compute $Z_{\textrm{\tiny eff}}$ as a weighted average of composing elements in the compound, {\it ii)} and methods that rely on mass attenuation coefficients of composing elements to compute $Z_{\textrm{\tiny eff}}$ \cite{bonnin_concept_2014}. The second group of methods was developed to solve the fundamental problem: $Z_{\textrm{\tiny eff}}$ as the quantity defined in the first category is not specifically tied to the absorption property of a material. Thus, weighted sums of elements in a compound used in the first category are exclusively valid for certain energy ranges and often to a certain set of elements contained in the mixture, while methods in the second category rely on tabulated attenuation properties of materials to inherently define $Z_{\textrm{\tiny eff}}$ as a quantity tied to attenuation property of the material.

In this section, we compared several available methods for the materials in the calibration phantom. Z\emph{comp}ARE \cite{vrbaski_zcompare_2022} is user-friendly software with a graphical interface that can be used to compare several methods most often used. The comparison of the methods for the calibration phantom materials in the energy range 20 - 40 keV is given in \ref{Tab5}. 

\begin{table}[ht]
\centering
\caption{Comparison of the most often used formulations of effective atomic numbers for phantom materials in the energy range 20 - 40 keV.\label{Tab5}}
\footnotesize
\begin{tabular}{@{}lllllll}
\br
\textbf{Material} & \textbf{Water} & \textbf{PE} & \textbf{PA} & \textbf{PMMA} & \textbf{POM} & \textbf{PTFE}\\
\mr
Brute formula & $\textrm{H}_2\textrm{O}$ & $\textrm{C}_2\textrm{H}_4$ & $\textrm{C}_{12}\textrm{H}_{22}\textrm{N}_2\textrm{O}_2$ & $\textrm{C}_5\textrm{H}_8\textrm{O}_2$ & $\textrm{C}\textrm{H}_2\textrm{O}$ & $\textrm{C}_2\textrm{F}_4$\\
\mr
Spiers {\it et al} & 7.42 & 5.44 & 6.12 & 6.47 & 6.95 & 8.43\\
Glasser {\it et al} & 7.96 & 5.94 & 6.60 & 6.94 & 7.38 & 8.62\\
Hine {\it et al} & 3.34 & 2.67 & 3.27 & 3.60 & 4.00 & 7.99\\
Puumalainen {\it et al} & 3.33 & 2.67 & 3.27 & 3.61 & 4.00 & 8.00\\
Tsai and Cho & 7.44 & 5.47 & 6.15 & 6.50 & 6.98 & 8.45\\
Gowda {\it et al} & 7.24 & 5.05 & 5.89 & 6.32 & 6.85 & 8.53\\
Champley {\it et al} & 7.44 & 5.28 & 6.16 & 6.49 & 7.01 & 8.56\\
\br
\end{tabular}
\end{table}
\normalsize

\begin{figure}[ht!]
\captionsetup[subfigure]{labelformat=empty}
\centering
 \begin{subfigure}{0.3\textwidth}
  \centering
  \includegraphics[width=\linewidth]{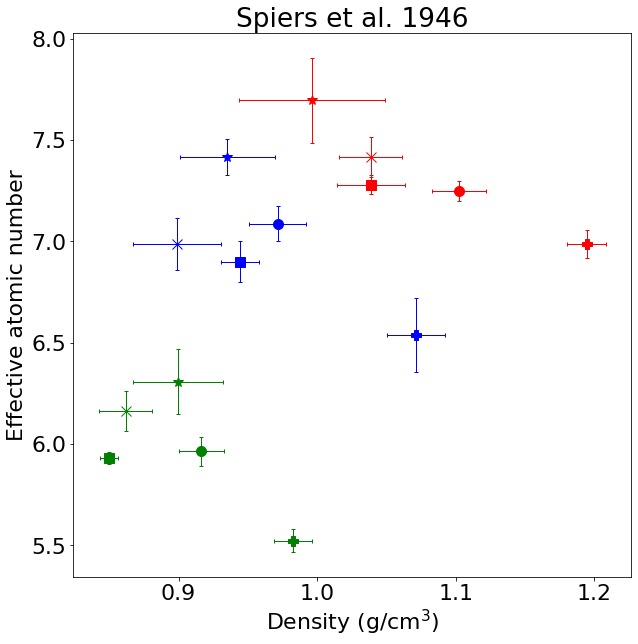}
  \caption{}
 \end{subfigure}
 \quad
 \begin{subfigure}{0.3\textwidth}
  \centering
  \includegraphics[width=\linewidth]{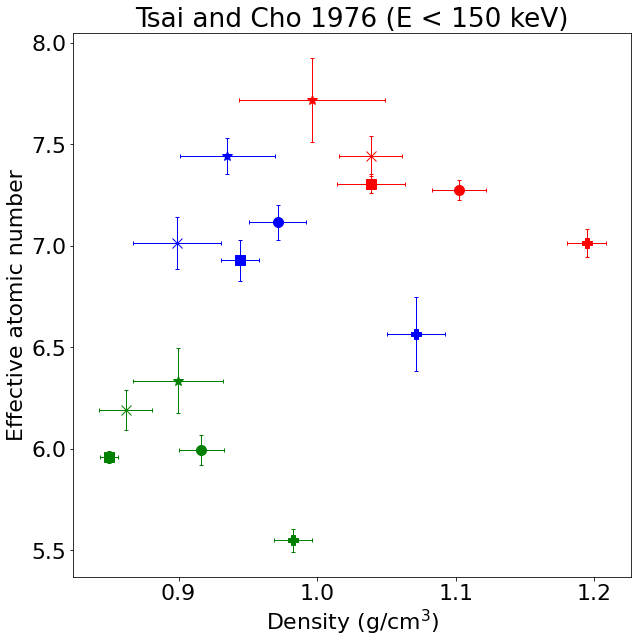}
  \caption{}
 \end{subfigure}
 \begin{subfigure}{0.3\textwidth}
  \centering
  \includegraphics[width=\linewidth]{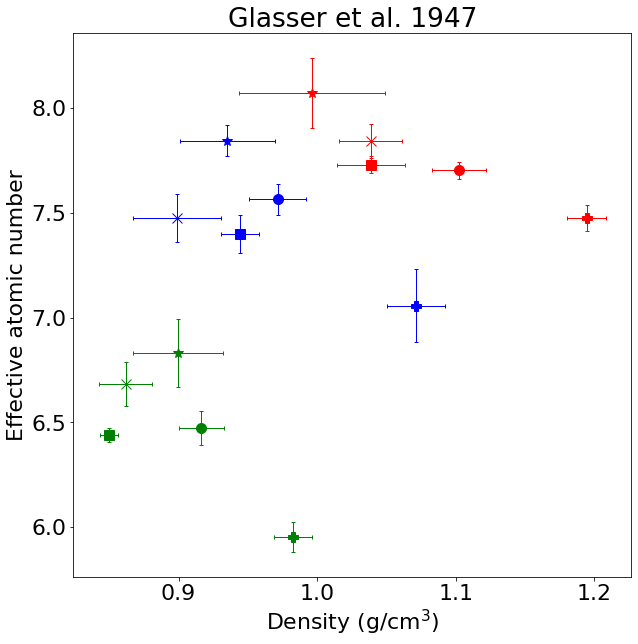}
  \caption{}
 \end{subfigure}
 \quad
 \begin{subfigure}{0.3\textwidth}
  \centering
  \includegraphics[width=\linewidth]{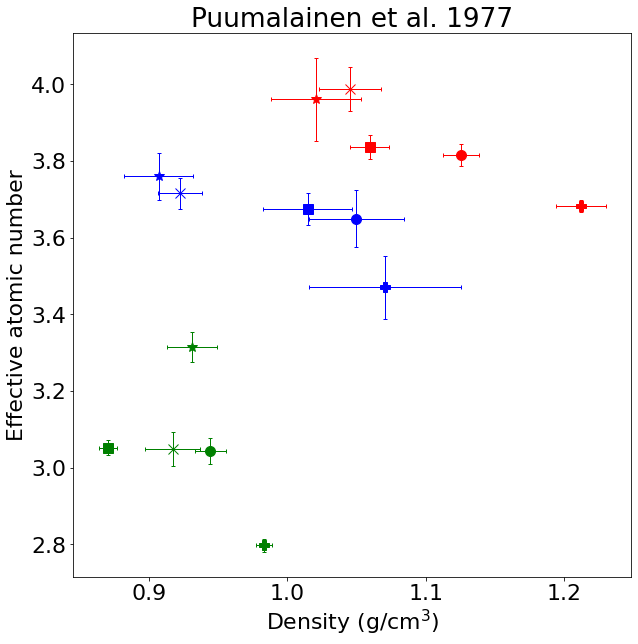}
  \caption{}
 \end{subfigure}
 \begin{subfigure}{0.3\textwidth}
  \centering
  \includegraphics[width=\linewidth]{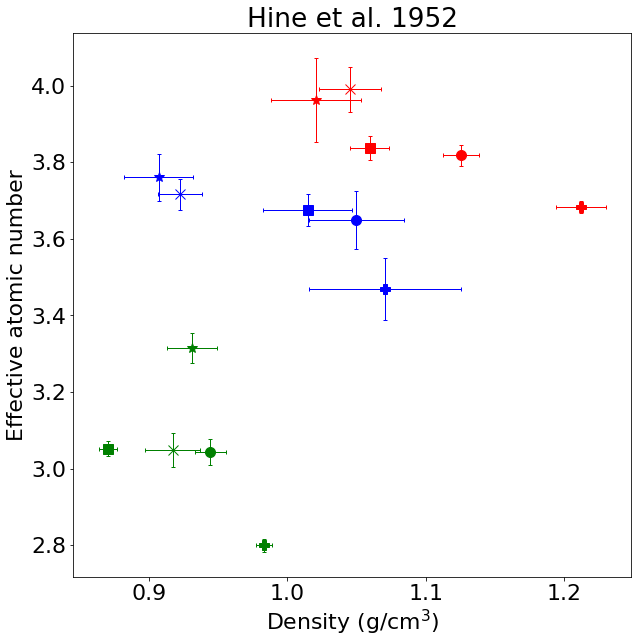}
  \caption{}
 \end{subfigure}
 \quad
 \begin{subfigure}{0.3\textwidth}
  \centering
  \includegraphics[width=\linewidth]{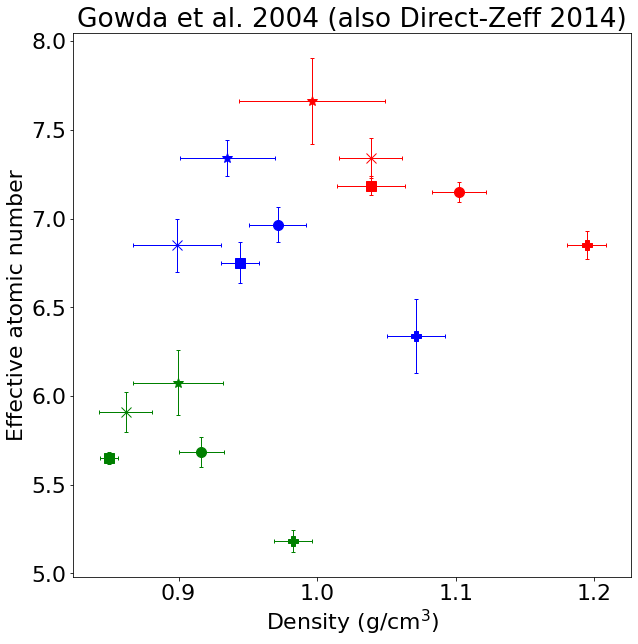}
  \caption{}
 \end{subfigure}
 \begin{subfigure}{0.45\textwidth}
  \centering
  \includegraphics[width=\linewidth]{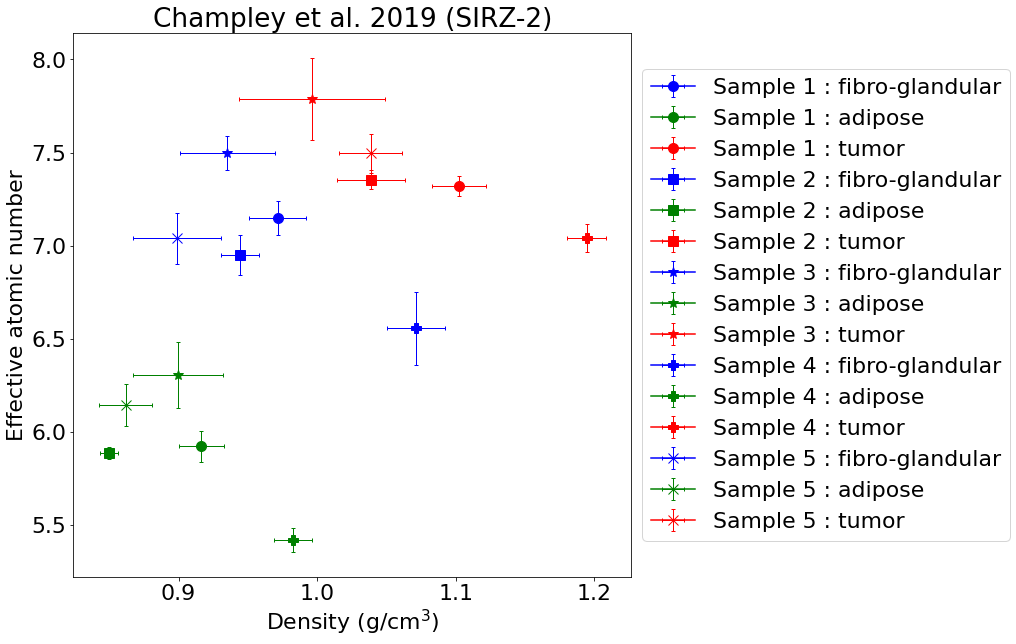}
  \caption{Comparison of $Z_{\textrm{\tiny eff}}$ definitions in $\rho$/$Z_{\textrm{\tiny eff}}$ space.}
 \end{subfigure}
\caption{}\label{fig:rhoZcompare}
\end{figure}

It can be seen that methods by Spiers {\it et al}, Tsai and Cho, Glasser {\it et al}, Gowda {\it et al}, and Champley {\it et al} provide very similar $Z_{\textrm{\tiny eff}}$ numbers for our experimental conditions. Methods by Hine {\it et al} and Puumalainen {\it et al} gave considerably lower $Z_{\textrm{\tiny eff}}$ numbers for all materials. Calibration functions of the form defined in equations (\ref{xrhofunc}) and (\ref{xZfunc}) were applied to all definitions for $Z_{\textrm{\tiny eff}}$ and results are given in figure \ref{fig:rhoZcompare}. The mean Silhouette score defining the level of separation in the range -1 to 1 was found to be approx. 0.31 for all methods used. Thus, the distinguishment of tissues could be obtained irrespective of the proposed definition. 

The method defined by Champley {\it et al} \citeyear{champley_method_2019} showed slightly better agreement of virtual monochromatic $\mu$ values computed using equation \ref{eq:mu} with experimentally measured data. Constants $n$, $K_1$, and $K_2$ used in equation \ref{eq:mu} were estimated to be 4.44, 9.4, and 1.6 using the least-square fit method for elemental materials found in the human body in the Z range of 1-20 across the diagnostic energy range (20 - 200 keV).

\pagebreak

\bibliographystyle{dcu.bst}

\section*{References}
\bibliography{arXiv_submission_breast_rho_Z}
\end{document}